# TOPAS-MC Extension for Nuclear Medicine Applications

Rodrigues, C.[1,2], Peralta, L.[1,2], Ferreira, P[3].

[1]Faculdade de Ciências, Universidade de Lisboa
[2]Laboratório de Instrumentação e Física Experimental de Partículas
[3]Champalimaud Centre for the Unknown, Champalimaud Foundation

**Abstract**

Monte Carlo (MC) techniques are currently deemed the gold standard for internal dosimetry, since the simulations can perform full radiation transport and reach a precision level not attainable by analytical methods. In this study, a custom voxelized particle source was developed for the TOPAS-MC toolkit to be used for internal dosimetry purposes. The source was designed to allow the use of clinical functional scans data to simulate events that reproduce the patient-specific tracer biodistribution. Simulation results are very promising, showing that this can be a first step towards the extension of TOPAS-MC to nuclear medicine applications. In the future more studies are needed to further ascertain the applicability and accuracy of the developed routines.

## I. Introduction

Over the last decade several methods have been used to quantify absorbed dose. With voxel-level dosimetry, one can obtain 3D dose maps, considering the non-uniform activity distribution within the patient's body, given by functional scans as in SPECT. Currently, Monte Carlo (MC) techniques are considered the gold standard, as they also allow the inclusion of patient-specific data about the tissue's heterogeneity, scattering and attenuation behavior, provided by structural images. Using sophisticated mathematical models, MC simulations can perform full radiation transport and reach a precision level not attainable by analytical methods [1][2]. Besides accounting for local energy deposition, MC simulations also consider voxel cross-irradiation, which is particularly important when studying the use of isotopes with γ-emission due to the longer path-length compared to α- and β-particles [3].

Multiple general-purpose MC codes are currently available, such as EGS4 [4], MCNP [5], PENELOPE [6] and Geant4 [7]. Nowadays, Geant4, a software toolkit originally created for high energy physics applications, is considered by many the main reference in the field of medical physics [8]. It is the most common choice for nuclear medicine applications due to its accurate physics models [9]. However, Geant4 requires C++ programming skills, which restricts its manipulation to specialists. Attempts were made to bring the sophisticated and reliable tools of Geant4 into experimentally validated and intuitive MC codes. GATE [10], TOPAS [11], and GAMOS [12] are examples. These programs are not generally implemented in a clinical setting, since they are computational demanding and time-consuming.

In this study, we investigated the use of the TOPAS (TOol for PArticle Simulation) MC toolkit for internal dosimetry purposes. To the best of our knowledge, this has never been reported. TOPAS is a breakthrough software project that became the benchmark for absorbed dose calculations in proton therapy. The radiobiology extension came years after its release in 2009 and, recently, a brachytherapy module was included. TOPAS overcomes the issues related with Geant4, by requiring little programming skills and providing great flexibility. The use of this toolkit, both in research centers and medical institutions, has been rapidly increasing as its advantages are recognized [11][13].

## II. Image-based Particle Source

A routine for TOPAS-MC was developed using Visual Studio Code v1.54.3 (Microsoft Corporation, United States), to create a custom particle source. This source was specifically designed for nuclear medicine applications (i.e., radionuclide-based diagnosis and therapy), to allow use of the functional scans (i.e., SPECT, PET) data to simulate events that reproduce the tracer biodistribution within the patient. To this date, TOPAS-MC did not support this feature.

The algorithm uses the functional data, given by the DICOM images, converted into a text file format to be imported to TOPAS, which must contain the size (in mm) and number of voxels in each dimension, and their intensity values. The rationale is that the quantitative data is used to define the distribution of the radiation source position in a proportional manner to the voxels' intensity (i.e., the number of events). The particle position is generated using the inverse transform method applied to a discrete distribution of stochastic variables, $x_i$ ($i = 1, …, N$), such as the number of arbitrary counts in the scans, with a probability distribution, $p_i$. The intensity data is read as a vector with $Nx \times Ny \times Nz$ elements (i.e., total number of elements in the original 3D matrices, where Nx, Ny and Nz are, respectively, the number of elements in x, y and z). The vector values are normalized to the total number of counts, and the cumulative probability discrete function is computed for each voxel (Eqs. (1) and (2)).

$$P_j = p_j + p_{j-1} + \cdots + p_1 \quad if \quad x_{j-1} \leq x < x_j \quad (1)$$

$$P_n = p_n + p_{n-1} + \cdots + p_1 = 1 \quad (2)$$

$$j, n = 1, …, Nx + Ny + Nz$$

where $x$ represents the voxel's unidimensional index. Using Geant4's default random number generator (HepJamesRandom engine), a random number ξ uniformly distributed in the interval (0,1) is generated (ξ → U(0,1)). This value (key element) is searched for in the cumulative probability vector, using the binary search algorithm. The index of the middle voxel of the array is selected as the partitioning element. This is used in a pairwise comparison made to check whether ξ is equal, higher (a) or lower (b) than the partitioning element.

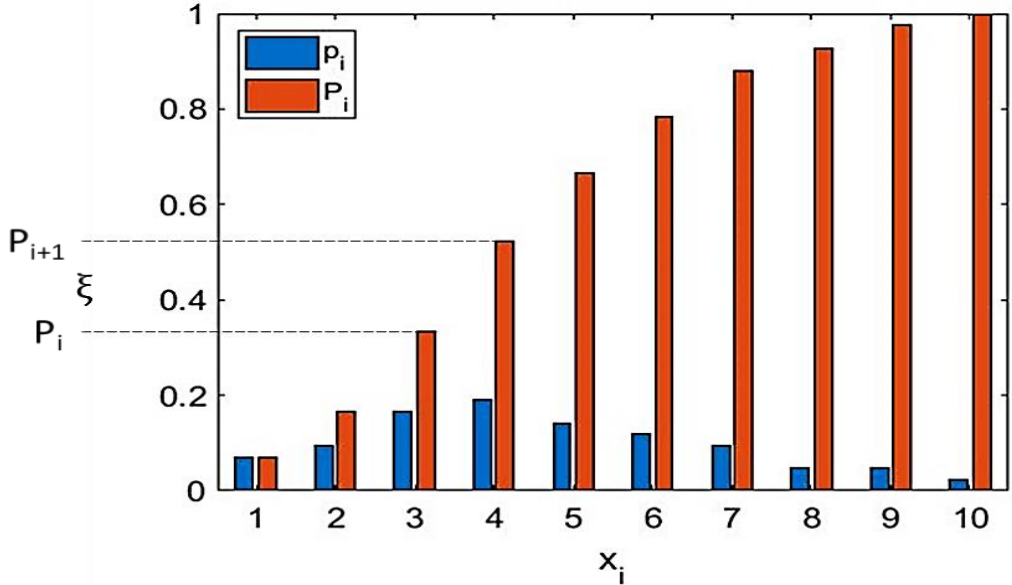

Figure 1. Example of the behavior of the implemented search algorithm to find the voxel in which the generated random number, ξ, is included. The integrated array values are represented in orange and the original array is shown in blue. If ξ has no direct correspondence in $x$, being between two values, it is considered to fall in the voxel of the immediately upper value.

If the values match, the position of the middle element in the array is returned. If the elements are not equal, the initial array is partitioned in two subarrays, with reference to the middle index. If condition (a) is verified, the lower half of the array is discarded. Contrary, if condition (b) is true, the upper half is rejected. The process is repeated for the remaining subarray and continues progressively until the key element, with the same value as ξ, is found. If ξ has no direct correspondence in the array, standing between two values, it is considered to fall in the voxel of the immediately upper value. The distribution of *x* is obtained through the condition

$$X = x_j \; if \; P_{j-1} < \xi \leq P_j, j \neq 1. \qquad (3)$$

Voxels with higher intensity values will be selected more often, such that the particles' position is sampled proportionally to the number of decays in the voxels with respect to the total number of events. The unidimensional index of the voxel whose value is equal to ξ corresponds to a voxel at a certain position in the original 3D image. The 3D coordinates are determined to obtain the spatial position of the particle. Within the selected voxel, the coordinates are sampled randomly and uniformly in each of the three dimensions. To avoid stuck particles in the voxels' boundaries, particularly when these define the transition between two different media, an offset can be considered so that the sampling region within the voxel is smaller than its actual extent. A new set of coordinates that defines the particle's position is generated in each event.

This extension supports the simultaneous spatial translation of the source volume (e.g., phantom or patient) and voxelized source. However, no implementations regarding rotation were included. To agree with TOPAS-MC placement of volumes in the spatial system, by default, the voxelized source is placed centered at the spatial system, being automatically aligned with the source volume (i.e., null translation in the parameter file for both volumes). This custom source is specified as any other source available in TOPAS-MC, in the main parameters file. However, its use necessarily requires that a text file with the intensity data is provided as input.

## III. Application Example

The particle source was tested by calculating absorbed dose maps using patient data on TOPAS-MC v3.6.1, based on Geant4 v10.06.p01. The code was compiled on a Linux Ubuntu 20 workstation equipped with Intel Core i7 9$^{th}$ Gen processors @ 2.6 GHz (16 GB RAM). No variance reduction techniques were applied. The complete radioactive decay of the emitting source was simulated using the G4RadioactiveDecayPhysics and G4Decay modules. Electromagnetic interactions were simulated using Geant4's standard EM physics package, which includes photoelectric effect, pair production, Compton and Rayleigh scattering for photon interactions, and bremsstrahlung, atomic ionization and multiple scattering for electron interactions. Fluorescence, PIXE and Auger electron emission was also enabled. A cutoff of 0.05 mm was set to all radiation and particles. The patient-specific CT scan was used define a voxelized anthropomorphic phantom that reproduces the actual patient anatomy (Figure 2) and TOPAS-MC built-in Hounsfield Units (HU) to material converter was used.

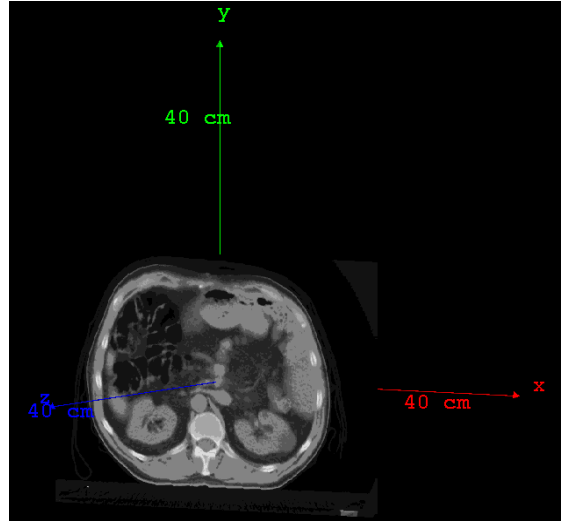

Figure 2. OpenGL visualization of the setup of a TOPAS-MC simulation. Only a single slice of the voxelized phantom is shown. CT materials are shown in different colors according to their HU value.

SPECT data was converted to text files in MATLAB and used to define the radiotracer's spatial distribution. Both the anthropomorphic phantom and radioactive source volumes were centered in the TOPAS spatial reference system. The particles (i.e., radionuclide used in imaging) were sampled randomly within the patient but according to the activity distribution. Simulations were performed using $10^8$ events. Each event corresponded to one full decay including multiple primary particles. The absorbed dose due to all primary and secondary particles was scored at the voxel-level on a grid overlaid on the voxelized phantom, using the HU-material conversion table. From the simulations we obtained 3D absorbed dose maps, covering all the CT scan volume (Figure 3). Each simulation took around 2 days to be completed.

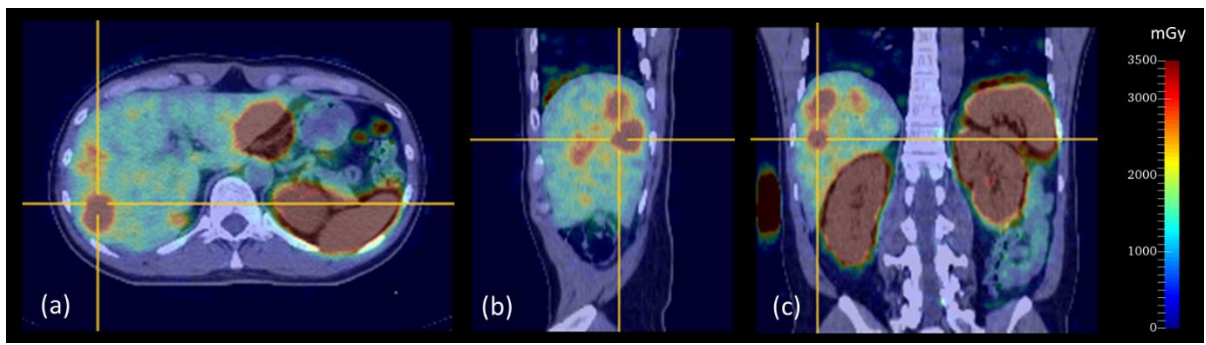

Figure 3. (a) axial, (b) sagittal and (c) coronal orthogonal projections of patient-specific activity-derived absorbed dose maps, aligned with the respective CT scan, obtained using TOPAS-MC simulations and the designed custom source. The color scale represents the absorbed dose in mGy. All projections are in the same scale. More absorbed dose is registered in the kidneys, spleen and tumors, which is in agreement with the radiotracer's distribution seen in the patient's SPECT scan.

## IV. Conclusions

In this preliminary study, new routines were developed for TOPAS-MC to enable patient-specific 3D dosimetry using clinical SPECT or PET/CT scans, which was not yet possible with the available features of the toolkit. The designed voxelized particle source enables the emission of particles from randomly and uniformly sampled starting positions within the source's volume, according to the distribution given by functional scans. The simulation outcomes were very promising. We believe this can be a first step towards the extension of TOPAS to nuclear medicine. However, to further ensure the proper behavior and applicability of this custom source, in the future, it should be validated in dosimetry calculations against similar well-validated tools of other MC codes (e.g., GATE).